\documentclass[aps,prl,twocolumn,groupedaddress]{revtex4}

\usepackage{amsmath}
\usepackage[dvips]{graphicx}

\begin{document}

%\preprint{}

\title{
Ferromagnetism of transition metals and screened exchange interactions
}



\author{Y. Kakehashi}
\email[]{yok@sci.u-ryukyu.ac.jp}
%\thanks{}
\affiliation{Department of Physics and Earth Sciences,
Faculty of Science, University of the Ryukyus,
1 Senbaru, Nishihara, Okinawa, 903-0213, Japan}

\author{M. Atiqur R. Patoary}
%\email[]{}
%\thanks{}
\affiliation{Department of Physics and Earth Sciences,
Faculty of Science, University of the Ryukyus,
1 Senbaru, Nishihara, Okinawa, 903-0213, Japan}


\date{\today}
\vspace*{10mm}
\begin{abstract}
We have investigated the magnetic properties of Fe, Co, and Ni at finite
temperatures on the basis of the first-principles dynamical coherent
potential approximation (CPA) in order to clarify the role of the
exchange interaction energy ($J$) screened by {\it sp} electrons and its
applicability to finite-temperature magnetism. 
With use of the atomic $J$, we obtained the Curie temperatures 
($T_{\rm C}$) 1930 K for Fe and 2550 K for fcc Co, which are
overestimated by a factor of 1.8 as compared with the experimental
values, while we obtained $T_{\rm C}=620$ K for Ni being in good 
agreement with
the experiment.  Calculated effective Bohr magneton numbers also
quantitatively agree with the experimental values.  By comparing the
results with those obtained by the screened $J$ and by comparing them 
with the experiments, 
we found that the screened $J$, which are reduced by 30\% as compared
 with the atomic ones, improve the ground-state
magnetizations and densities of states at low temperatures in Fe and
fcc Co, as well as $T_{\rm C}$ in fcc Co.  But the screened $J$ yield
worse results for finite temperature properties of Fe and underestimate
both the ground-state magnetization and $T_{\rm C}$ in case of Ni.
We discuss possible origins for these inconsistencies.
\end{abstract}

%\keywords{}

\maketitle


\section{Introduction}

%\subsection{}
%\subsubsection{}

The band theory of itinerant magnetism has been much
developed in the past half a century on the basis of the density
functional theory (DFT)~\cite{hohen64,kohn65,parr89} 
with local spin density approximation (LSDA)~\cite{barth72} or 
the generalized gradient approximation (GGA)~\cite{perdew86}.  
The DFT explains
quantitatively the ground-state magnetism of 3$d$ transition metals and
alloys. The first-principles band calculations with use of the LSDA, for
example, yield the magnetization per atom 2.15 $\mu_{\rm B}$ for Fe 
and 0.59 $\mu_{\rm B}$ for Ni,
which are in good agreement with the experimental 
values~\cite{danan68,bes70} 2.22 
$\mu_{\rm B}$ and 0.62 $\mu_{\rm B}$, respectively.  The LSDA stabilizes
artificially the fcc Fe instead of the bcc Fe with the volume
contraction by 15 \%. The GGA has solved the problem on the stability of 
structure in Fe~\cite{bagno89}.
 
On the other hand, understanding of the finite-temperature magnetism in
transition metals is still far from the final goal of quantitative
description.  The Stoner theory based on the DFT band calculations
yields the Curie temperatures ($T_{\rm C}$), 6000 K for Fe and 3000 K
for Ni~\cite{gunnarsson77,staunton92}, 
which are 6 or 5 times as large as the observed values (1040 K
for Fe~\cite{arrott67} and 630 K for Ni~\cite{noakes66}).  
Because of a large discrepancy between the
theory and the experiment, many theories which take into account spin
fluctuations at finite temperatures have been developed~\cite{kake04}.  
Hubbard~\cite{hub79} and Hasegawa~\cite{hase79} proposed a
single-site spin fluctuation theory (SSF) on the basis of the functional
integral method~\cite{strat58,hub59,evan70,mora74} 
and the coherent potential approximation (CPA)~\cite{soven67,ehren76}.
They explained the Curie-Weiss susceptibility in 3$d$ transition metals
and obtained the Curie temperatures in Fe and Ni which qualitatively 
agree with the experiments.

The SSF, however, reduces to the Hartree-Fock theory at $T=0$ because it
is based on a high-temperature approximation.  Therefore, the theory
does not take into account electron correlations at the ground state 
as discussed by Gutzwiller~\cite{gutz63,gutz64,gutz65}, 
Hubbard~\cite{hub63,hub64}, and Kanamori~\cite{kana63}.
Kakehashi and Fulde~\cite{kake85} 
proposed a variational theory at finite temperatures
which takes into account the Gutzwiller-type electron correlations at
the ground state and reduces to the SSF at high temperatures.  
They found that the
Curie temperatures in Fe and Ni are reduced by $30\% \sim 50\%$ due to 
correlated motion of electrons.  
Hasegawa~\cite{hase90} also proposed a similar theory on the
basis of the slave-boson functional integral method.
These theories are not suitable for systematic improvements of the
theory because of the difficulty in finding a suitable temperature
dependence of the effective potential being projected onto the static
fields.  In order to make a systematic improvements possible, we
proposed the dynamical CPA~\cite{kake92} which fully takes into account the
single-site spin and charge fluctuations self-consistently, and
clarified the basic properties of the theory with use of a Monte-Carlo
method.  In the next paper~\cite{kake02} which we refer to I, 
we proposed more analytic method to the dynamical CPA 
using the harmonic approximation (HA)~\cite{amit71,dai91}.

Towards quantitative calculations, it is indispensable to take into
account realistic band structures in solids.  
In our recent paper~\cite{kake08} which
we refer to II, we proposed the first-principles dynamical CPA 
combining the dynamical CPA+HA with the LDA+U scheme~\cite{anis97} 
based on the tight-binding linear muffin-tin orbitals 
(TB-LMTO)~\cite{ander94}. 
Quite recently we extended the calculations taking into
account the dynamical corrections up to the 4th order in Coulomb
interactions, and have shown that the single-particle excitation 
spectra obtained by the first-principles dynamical CPA explain
systematic change of the X-ray photoemission spectroscopy (XPS) and 
the bremsstrahlung isochromat (BIS) data from Sc to Cu at 
high-temperatures~\cite{kake09,kake10}.

In our numerical calculations, we adopted the Coulomb interactions
$U$ obtained by Anisimov {\it et. al.}~\cite{anis97-2}, 
and by Bandyopadhyay and Sarma~\cite{bdyo89}. 
The exchange interaction energies $J$ were taken from the values
obtained by the Hartree-Fock atomic calculations~\cite{mann67}.  
Recently, considerable efforts to obtain 
the first-principles $U$ and $J$ have been made at the ground state. 
Aryasetiawan and coworkers~\cite{arya06} obtained the Coulomb and
exchange interactions for $d$ orbitals which are screened by $sp$
electrons, using the random phase
approximation (RPA) and the LMTO method.  
The same type of calculations has been
performed by using the first-principles Wannier orbitals by Miyake and
Aryasetiawan~\cite{miyake08}. 
These results indicate that irrespective of transition
metal elements the exchange energy integral $J$ are reduced by about 
$30\%$ as compared with the atomic ones.  
Although the methods have been applied to 
many other compounds~\cite{imada10}, there 
is no systematic investigation for the influence of the screened $J$ on 
magnetic properties. 

We investigate in this paper the ferromagnetic properties of Fe, Co, 
and Ni on the basis of the first-principle dynamical
CPA using both the atomic exchange interaction energies (atomic $J$) and 
the exchange interaction energies screened by 30\% (screened $J$).  
We examine the role of the exchange energy parameters
in the magnetic properties at finite temperatures and clarify
their validity to the magnetic properties.
We will demonstrate that use of the screened $J$ 
does not necessary improve theoretical results at finite temperatures,
though we find some improvements at low temperature regime.

As we have proven in our previous papers~\cite{kake04,kake02-2}, 
the dynamical CPA is equivalent to the many-body CPA~\cite{hiro77} in
disordered alloys, the dynamical mean-field theory 
(DMFT)~\cite{mull89,jarr92,ohkawa92,geor93,geor96} 
in the metal-insulator transitions,
and the projection operator CPA~\cite{kake04-1} 
for excitation problem in solids.  The first-principles DMFT calculations 
for Fe and Ni have been performed at the ground state by Miura and 
Fujiwara~\cite{miura08} 
within the iterative perturbation method.
The finite-temperature DMFT calculations for Fe and Ni
have been performed by using the Hamiltonian without transverse spin
fluctuations~\cite{lich01}.  
In both cases, atomic exchange interaction energy $J=0.066$ Ry was
adopted for Fe and Ni.  
We present here the finite-temperature results obtained by 
the Hamiltonian with transverse spin fluctuations and by using both the
atomic $J$ and the screened $J$ values.

The outline of the paper is as follows.  In the following section, we
summarize the first-principles dynamical CPA based on the HA.
In Sec. III, we present numerical results of calculations for 
the densities of states as
the excitation spectra, magnetization vs temperature curves, Curie
temperatures, paramagnetic susceptibilities, effective Bohr
magneton numbers, as well as the amplitudes of local magnetic moments in
Fe, Co, and Ni.  By comparing these quantities calculated by the atomic
$J$ with those obtained by the screened $J$, we will clarify the role of
the screening effects on the exchange energy $J$, and discuss the 
availability of the screened $J$.  
Furthermore, we discuss the validity of the screened $J$ by comparing
calculated magnetic properties with those in experiments.
In Sec. IV, we present a summary of 
our works, and discuss possible reasons for the disagreement with 
experiments when the screened $J$ are applied.

\section{First-principles dynamical CPA}

In the first-principles dynamical CPA, we adopt the TB-LMTO Hamiltonian 
combined with a LDA+U Coulomb interactions as follows~\cite{kake08}.  
\begin{eqnarray}
H = H_{0} + H_{1} ,
\label{hhat}
\end{eqnarray}
\begin{eqnarray}
H_{0} = \sum_{iL\sigma} (\epsilon^{0}_{L} - \mu) \, \hat{n}_{iL \sigma} 
+ \sum_{iL jL^{\prime} \sigma} t_{iL jL^{\prime}} \, 
a_{iL \sigma}^{\dagger} a_{jL^{\prime} \sigma} \ ,
\label{h0}
\end{eqnarray}
\begin{eqnarray}
H_{1} &=& \sum_{i} 
\Big[ \sum_{m} U_{0} \, \hat{n}_{ilm \uparrow} \hat{n}_{ilm \downarrow} 
+ {\sum_{m > m^{\prime}}} 
(U_{1}-\frac{1}{2}J) \hat{n}_{ilm} \hat{n}_{ilm^{\prime}} \nonumber\\ 
& &-{\sum_{m > m^{\prime}}} J   
\hat{\mbox{\boldmath$s$}}_{ilm} \cdot \hat{\mbox{\boldmath$s$}}_{ilm^{\prime}} 
\Big] \ . 
\label{h1}
\end{eqnarray}
We assumed here a transition metal with an atom per unit cell.
$\epsilon^{0}_{L}$ in Eq. (\ref{h0}) is an atomic level on site $i$ and 
orbital $L$, $\mu$ is the chemical potential, 
$t_{iL jL^{\prime}}$ is a transfer integral between orbitals $iL$ and 
$jL^{\prime}$. $L=(l,m)$ denotes $s$, $p$, and $d$ orbitals.
$a_{iL \sigma}^{\dagger}$ 
($a_{iL \sigma}$) is the creation (annihilation) operator for an
electron with orbital $L$ and spin $\sigma$ on site $i$, and 
$\hat{n}_{iL\sigma}=a_{iL \sigma}^{\dagger}a_{iL \sigma}$ is a charge
density operator for electrons with orbital $L$ and spin $\sigma$ on
site $i$. 
 
The Coulomb interaction term $H_{1}$ in Eq. (\ref{h1})
consists of the on-site interactions between $d$ electrons ($l=2$).
$U_{0}$ ($U_{1}$) and $J$ denote the intra-orbital (inter-orbital)
Coulomb and exchange interactions, respectively.  
$\hat{n}_{ilm}$ ($\hat{\mbox{\boldmath$s$}}_{ilm}$) with $l=2$ is 
the charge (spin)
density operator for $d$ electrons on site $i$ and orbital $m$.
Note that the atomic level $\epsilon^{0}_{L}$ in 
$H_{0}$ is not identical with the LDA atomic level
$\epsilon_{L}$; $\epsilon^{0}_{L} = \epsilon_{L} - \partial
E^{U}_{\rm LDA}/\partial n_{iL\sigma}$. 
Here $n_{iL\sigma}$ is the charge density at the ground state, 
$E^{U}_{\rm LDA}$ is a LDA functional to the intra-atomic Coulomb 
interactions~\cite{anis97,anis97-2}.

In the dynamical CPA, we transform in the free energy the interacting 
Hamiltonian $H_{1}$
into a one-body Hamiltonian with dynamical potential $v$ for 
time-dependent random
charge and exchange fields, using the
functional integral method~\cite{mora74,kake08}.  
Introducing a site-diagonal uniform medium, 
{\it i.e.}, a coherent potential $\Sigma$ into the potential part, 
we expand the correction $v-\Sigma$ with respect
to sites in the free energy.  
The zeroth term in the expansion is the free energy for 
a uniform medium, $\tilde{\cal F}[\Sigma]$.  
The next term is an impurity contribution to the
free energy.  The dynamical CPA neglects the higher-order terms
associated with inter-site correlations.
The free energy per atom is then given by~\cite{kake02,kake08}
\begin{eqnarray}
{\mathcal F}_{\rm CPA} = \tilde{\mathcal F}[\Sigma]
- \beta^{-1} {\rm ln} \, \int \Big[ \prod_{\alpha} 
\sqrt{\dfrac{\beta \tilde{J}_{\alpha}}{4\pi}}
d \xi_{\alpha} \Big] \,
{\rm e}^{\displaystyle -\beta E_{\rm eff}(\mbox{\boldmath$\xi$})} .
\label{fcpa2}
\end{eqnarray}
Here $\beta$ is the inverse temperature,  
$\tilde{J}_{x}=\tilde{J}_{y}=\tilde{J}_{\bot}=[1-1/(2l+1)]J$, 
$\tilde{J}_{z} = U_{0}/(2l+1) + \tilde{J}_{\bot}$, and
$\mbox{\boldmath$\xi$} = (\xi_{x}, \xi_{y}, \xi_{z})$ is a static 
field variable on a site.  

The effective potential 
$E_{\rm eff}(\mbox{\boldmath$\xi$})$ in Eq. (\ref{fcpa2}) consists of 
the static term $E_{\rm st}(\mbox{\boldmath$\xi$})$ and 
the dynamical correction term 
$E_{\rm dyn}(\mbox{\boldmath$\xi$})$ as follows.
\begin{eqnarray}
E_{\rm eff}(\mbox{\boldmath$\xi$}) = E_{\rm st}(\mbox{\boldmath$\xi$}) 
+ E_{\rm dyn}(\mbox{\boldmath$\xi$}) .
\label{eeff}
\end{eqnarray}
The static term is given as
\begin{widetext}
\begin{eqnarray}
E_{\rm st}(\boldsymbol{\xi}) &=& 
- \dfrac{1}{\beta} \sum_{mn} 
{\rm ln} \Big[ 
(1 \! - \! \delta v_{L\uparrow}(0)F_{L\uparrow}(i\omega_{n}))
(1 \! - \! \delta v_{L\downarrow}(0)F_{L\downarrow}(i\omega_{n}))
- \dfrac{1}{4} \tilde{J}^{2}_{\bot} \xi^{2}_{\bot} 
F_{L\uparrow}(i\omega_{n})F_{L\downarrow}(i\omega_{n})
\Big]     \nonumber \\
&  & 
+ \dfrac{1}{4} \Big[
- (U_{0}-2U_{1}+J) \sum_{m} \tilde{n}_{L}(\boldsymbol{\xi})^{2}
- (2U_{1}-J) \tilde{n}_{l}(\boldsymbol{\xi})^{2}
+ \tilde{J}^{2}_{\bot} \xi^{2}_{\bot} + \tilde{J}^{2}_{z} \xi^{2}_{z}
\Big] .
\label{est2}
\end{eqnarray}
\end{widetext}
Here 
$\delta v_{L\sigma}(0) = v_{L\sigma}(0) - \Sigma_{L\sigma}(i\omega_{n})$, 
and $\xi^{2}_{\bot}= \xi^{2}_{x} + \xi^{2}_{y}$.  
$v_{L\sigma}(0)$ is a static potential given by 
$v_{L\sigma}(0) = [(U_{0}-2U_{1}+J)\tilde{n}_{lm}(\boldsymbol{\xi})+
(2U_{1}-J)\tilde{n}_{l}(\boldsymbol{\xi})]/2 -
\tilde{J}_{z}\xi_{z}\sigma/2$,
$\Sigma_{L\sigma}(i\omega_{n})$ is the coherent potential for 
Matsubara frequency $\omega_{n}=(2n+1)\pi/\beta$.
The electron number $\tilde{n}_{L}(\boldsymbol{\xi})$ for a given 
$\boldsymbol{\xi}$ is expressed by means of an impurity Green function 
as
\begin{eqnarray}
\tilde{n}_{L}(\boldsymbol{\xi}) = \frac{1}{\beta} \sum_{n\sigma}
 G_{L\sigma}(\boldsymbol{\xi}, i\omega_{n}) ,
\label{nlxi}
\end{eqnarray}
and $\tilde{n}_{l}(\boldsymbol{\xi}) = 
\sum_{m} \tilde{n}_{L}(\boldsymbol{\xi})$.
The impurity Green function 
$G_{L\sigma}(\boldsymbol{\xi}, i\omega_{n})$ 
has to be determined self-consistently.  The explicit expression will 
be given later (see Eq. (\ref{gimp})). 
 
The coherent Green function $F_{L\sigma}(i\omega_{n})$ in Eq. (\ref{est2})
is defined by
\begin{eqnarray}
F_{L\sigma}(i\omega_{n}) = [(i\omega_{n} - \mbox{\boldmath$H$}_{0} 
- \mbox{\boldmath$\Sigma$}(i\omega_{n}))^{-1}]_{iL\sigma iL\sigma} .
\label{fls}
\end{eqnarray}
Here $(\mbox{\boldmath$H$}_{0})_{iL\sigma jL^{\prime}\sigma}$ 
is the one-electron Hamiltonian matrix for the noninteracting
Hamiltonian $H_{0}$, and 
$(\mbox{\boldmath$\Sigma$}(i\omega_{n}))_{iL\sigma jL^{\prime}\sigma} = 
\Sigma_{L\sigma}(i\omega_{n})\delta_{ij}\delta_{LL^{\prime}}$.

The dynamical potential $E_{\rm dyn}(\mbox{\boldmath$\xi$})$
in Eq. (\ref{eeff}) has been obtained within the harmonic
approximation (HA)~\cite{kake02,kake08,amit71,dai91}. 
It is based on an expansion of $E_{\rm dyn}(\boldsymbol{\xi})$ with
respect to the frequency mode of the dynamical potential 
$v_{L\sigma\sigma^{\prime}}(i\omega_{\nu})$, where 
$\omega_{\nu}=2\nu\pi/\beta$.  The HA is the neglect
of the mode-mode coupling terms in the expansion.  We have then  
\begin{eqnarray}
E_{\rm dyn}(\boldsymbol{\xi}) = - \beta^{-1} {\rm ln} 
\left[ 1 + \sum^{\infty}_{\nu=1} \,(\overline{D}_{\nu} -1) \right] .
\label{edyn1}
\end{eqnarray}
Here the determinant $D_{\nu}$ is a contribution from a dynamical
potential $v_{L\sigma\sigma^{\prime}}(i\omega_{\nu})$ with frequency
$\omega_{\nu}$, and the upper bar denotes a Gaussian
average with respect to the dynamical charge and exchange field
variables, $\zeta_{m}(i\omega_{n})$ and 
$\xi_{m\alpha}(i\omega_{n})$ ($\alpha = x, y, z$).  

The determinant $D_{\nu}$ is expressed as~\cite{kake08} 
\begin{eqnarray}
D_{\nu} = \prod_{k=0}^{\nu-1} \left[ \prod_{m=1}^{2l+1} D_{\nu}(k,m)
\right] ,
\label{dnu2}
\end{eqnarray}
%
%
%\begin{eqnarray}
%D_{\nu}(k,m) = \left| 
%\begin{array}{@{\,}ccccccc@{\,}}
%\ddots &         &         &      &    &   &  \\
%       & 1       & 1           &        & 0  & &  \\
%       & a_{-\nu+k}(\nu,m) & 1            & 1    &    & &  \\
%       &                & a_{k}(\nu,m) & 1        & 1  & &  \\
%       &                &      & a_{\nu+k}(\nu,m) & 1  & 1 & \\
%       & 0       &      &          & a_{2\nu+k}(\nu,m) & & \\
%       &                &      &          &                & \ddots \ \ \  & \\
%\end{array}
%\right| \ .
%\label{dnukm}
%\end{eqnarray}
%
\begin{eqnarray}
D_{\!\nu}(\!k,\!m\!) \!\!= \!\!\left| 
\begin{array}{@{\,}ccccccc@{\,}}
\ddots\!\!\!\!\!\!\!\!\!\!\!\!\! &         &         &      &    &   &  \\
       & 1       & 1           &        & 0  & &  \\
       & a_{\!-\!\nu\!+\!k}(\!\nu,\!m\!)\!\!\!\!\! & 1    & 1    &    & &  \\
       &               & a_{k}\!(\!\nu,\!m)\!\!\!\!\!\!\!\! & 1   & 1  & &  \\
       &            &      & a_{\nu\!+\!k}(\!\nu,\!m)\!\!\!\!\!\!\!\!\!\!\! & 1  & 1 & \\
       & 0       &      &       & a_{2\nu\!+\!k}(\!\nu,\!m)\!\!\!\!\!\! & & \\
       &          &      &          &                & \!\!\!\ddots \!\!\!\!\!\!\!& \\
\end{array}
\right| . \hspace{-0mm}
\label{dnukm}
\end{eqnarray}

Note that 1 in the above determinant denotes the $2 \times 2$ unit matrix, 
$a_{n}(\nu,m)$ is a $2 \times 2$ matrix in the spin space, which are
defined by 
\begin{eqnarray}
a_{n}(\nu,m)_{\sigma\sigma{\prime}} &=& 
\sum_{\sigma^{\prime\prime}\sigma^{\prime\prime\prime}
\sigma^{\prime\prime\prime\prime}} 
v_{L\sigma\sigma^{\prime\prime}}(i\omega_{\nu}) 
\tilde{g}_{L\sigma^{\prime\prime} \sigma^{\prime\prime\prime}}
(i\omega_{n}-i\omega_{\nu})\nonumber\\
& &\times v_{L\sigma^{\prime\prime\prime}\sigma^{\prime\prime\prime\prime}}
(-i\omega_{\nu}) 
\tilde{g}_{L\sigma^{\prime\prime\prime\prime} \sigma^{\prime}}(i\omega_{n}) \ ,
\label{annum}
\end{eqnarray}
\begin{eqnarray}
v_{L\sigma\sigma^{\prime}}(i\omega_{\nu}) &=& 
- \frac{1}{2}  \sum_{m^{\prime}} i A_{mm^{\prime}}
\zeta_{m^{\prime}}(i\omega_{\nu}) \delta_{l2}\delta_{\sigma\sigma^{\prime}} \nonumber \\
&&- \frac{1}{2} \sum_{\alpha} \sum_{m^{\prime}} 
B^{\alpha}_{mm^{\prime}} \xi_{m^{\prime}\alpha}(i\omega_{\nu})
\delta_{l2} 
(\sigma_{\alpha})_{\sigma\sigma^{\prime}} \ ,\nonumber\\
\label{dpot2}
\end{eqnarray}
\begin{eqnarray}
\tilde{g}_{L\sigma\sigma^{\prime}}(i\omega_{n}) = [(F_{L}(i\omega_{n})^{-1} - 
\delta v_{0})^{-1}]_{\sigma\sigma^{\prime}} \ .
\label{gst}
\end{eqnarray}
Here $\sigma_{\alpha}$ ($\alpha=x, y, z$) are the Pauli spin matrices.
$A_{mm^{\prime}} = U_{0}\delta_{mm^{\prime}} 
+ (2U_{1} - J)(1 - \delta_{mm^{\prime}})$, 
$B^{\alpha}_{mm^{\prime}} = J (1 - \delta_{mm^{\prime}})$ 
$(\alpha = x,y$), and $B^{z}_{mm^{\prime}} 
=  U_{0} \delta_{mm^{\prime}} + J (1 - \delta_{mm^{\prime}})$. 
$\tilde{g}_{L\sigma \sigma^{\prime}}(i\omega_{n})$ is the impurity
Green function in the static approximation,
$(F_{L}(i\omega_{n}))_{\sigma\sigma^{\prime}} = 
F_{L\sigma}(i\omega_{n})\delta_{\sigma\sigma^{\prime}}$, and 
$\delta v_{0}$ is defined by 
$(\delta v_{0})_{\sigma\sigma^{\prime}} = 
v_{L\sigma\sigma^{\prime}}(0) - 
\Sigma_{L\sigma}(i\omega_{n})\delta_{\sigma\sigma^{\prime}}$.

The determinant $D_{\nu}(k,m)$ defined by Eq. (\ref{dnukm})  
is expanded with respect to the dynamical potential as follows. 
\begin{eqnarray}
D_{\nu}(k,m) = 1 + D^{(1)}_{\nu}(k,m) + D^{(2)}_{\nu}(k,m) + \cdots ,
\label{dnukm2}
\end{eqnarray}
\begin{eqnarray}
D^{(n)}_{\nu}(k,m) &=& \sum_{\alpha_{1}\gamma_{1} \cdots \alpha_{n}\gamma_{n}}
v_{\alpha_{1}}(\nu,m)v_{\gamma_{1}}(-\nu,m) \cdots \nonumber\\
& &\times v_{\alpha_{n}}(\nu,m)v_{\gamma_{n}}(-\nu,m) 
\hat{D}^{(n)}_{\{ \alpha\gamma \}}(\nu,k,m) \ .\nonumber\\
\label{dnnukm}
\end{eqnarray}
Here the subscripts 
$\alpha_{i}$ and $\gamma_{i}$ take 4 values $0$, $x$, $y$, and $z$,
and
\begin{eqnarray}
v_{0}(\nu,m) = - \dfrac{1}{2} i \sum_{m^{\prime}} A_{mm^{\prime}}
\zeta_{m^{\prime}}(i\omega_{\nu})\delta_{l2} \ ,
\label{v0num}
\end{eqnarray}
\begin{eqnarray}
v_{\alpha}(\nu,m) = - \dfrac{1}{2} \sum_{m^{\prime}} 
B^{\alpha}_{mm^{\prime}} \xi_{m^{\prime}\alpha}(i\omega_{\nu})\delta_{l2} \ , 
\hspace*{3mm} (\alpha=x,y,z) \ .\nonumber \hspace*{-10mm}\\
\label{vanum}
\end{eqnarray}
Note that the subscript 
$\{ \alpha\gamma \}$ of $\hat{D}^{(n)}_{\{ \alpha\gamma \}}(\nu,k,m)$ 
in Eq. (\ref{dnnukm}) denotes a set of 
$(\alpha_{1}\gamma_{1}, \cdots, \alpha_{n}\gamma_{n})$.

Substituting Eq. (\ref{dnukm2}) into Eq. (\ref{dnu2}) and taking the
Gaussian average, we reach
\begin{eqnarray}
E_{\rm dyn}(\boldsymbol{\xi}) = - \beta^{-1} {\rm ln} 
\left( 1 + \sum^{\infty}_{n=1} \sum^{\infty}_{\nu=1} 
\overline{D}_{\nu}^{(n)} \right) ,
\label{edyn2}
\end{eqnarray}
and
\begin{eqnarray}
\overline{D}^{(n)}_{\nu} = \dfrac{1}{(2\beta)^{n}} 
\sum_{\sum_{km} l(k,m)=n} \sum_{\{ \alpha_{j}(k,m)\} }
\sum_{\rm P}
\prod_{m=1}^{2l+1} \prod_{k=0}^{\nu-1}\nonumber\\
\times \Bigg[ \Big( \prod_{j=1}^{l(k,m)} C^{\alpha_{j}(k,m)}_{mm_{\rm p}} \Big)
\hat{D}^{(l(k,m))}_{\{ \alpha\alpha_{{\rm p}^{-1}} \} }(\nu,k,m) \Bigg] .
\label{dnubarn}
\end{eqnarray}
Here each element of 
$\{ l(k,m)\}\, (k=0, \cdots, \nu-1, m=1, \cdots , 2l+1)$ has a value of 
zero or positive integer.
$\alpha_{j}(k,m)$ takes one of 4 cases $0$, $x$, $y$, and $z$.
$j$ denotes the $j$-th member
of the $(k,m)$ block with $l(k,m)$ elements. 
P denotes a permutation of a set $\{ (j,k,m) \}$; 
${\rm P} \{ (j,k,m) \} = \{ (j_{\rm p},k_{\rm p},m_{\rm p}) \}$.
$\alpha_{{\rm p}^{-1}}$ 
means a rearrangement of $\{ \alpha_{j}(k,m) \}$ according to 
the inverse permutation P${}^{-1}$.  
The coefficient $C^{\alpha}_{mm^{\prime}}$ in Eq. (\ref{dnubarn}) 
is a Coulomb interaction defined by 
\begin{eqnarray}
C^{\alpha}_{mm^{\prime}} = \begin{cases}
-A_{mm^{\prime}} & (\alpha=0) \\
B^{\alpha}_{mm^{\prime}}  & (\alpha=x,y,z) \ .
\end{cases}
\label{cdef}
\end{eqnarray}
The frequency dependent factors 
$\hat{D}^{(n)}_{\{ \alpha\gamma \}}(\nu,k,m)$ in Eqs. (\ref{dnnukm}) and 
(\ref{dnubarn})  
consist of a linear combination of $2n$ products of the static Green
functions.  Their first few terms have been given in Appendix A of 
our paper II~\cite{kake08}.

In the calculations of the higher-order dynamical corrections~\cite{kake10} 
$\hat{D}^{(n)}_{\{ \alpha\gamma \}}(\nu,k,m)$, 
we note that the coupling constants 
$B^{x}_{mm^{\prime}}=B^{y}_{mm^{\prime}}=J(1-\delta_{mm^{\prime}})$
are considerably smaller than $A_{mm^{\prime}}$ and $B^{z}_{mm^{\prime}}$
because $U_{0}$ and $U_{1} \gg J$.  Thus we neglect the transverse
potentials, $v_{x}(\nu, m)$ and $v_{y}(\nu, m)$. 
The approximation implies that $a_{n}(\nu, m)_{\sigma -\sigma}=0$.  
The determinant $D_{\nu}(k,m)$ in Eq. (\ref{dnu2}) is then written 
by the products of the single-spin components as
\begin{eqnarray}
D_{\nu}(k,m) = D_{\nu\uparrow}(k,m)D_{\nu\downarrow}(k,m) .
\label{dnu3}
\end{eqnarray}
Here $D_{\nu\sigma}(k,m)$ is defined by Eq. (\ref{dnukm}) in which 
the $2 \times 2$ unit matrices have been replaced by 1 ({\it i.e.}, 
$1 \times 1$ unit matrices), and the $2 \times 2$ matrices 
$a_{n}(\nu, m)$ have been replaced by the $1 \times 1$ matrices 
$a_{n}(\nu, m)_{\sigma\sigma}$.  The latter is given by
\begin{eqnarray}
a_{n}(\nu, m)_{\sigma\sigma} = \sum^{0,z}_{\alpha,\gamma}
 v_{\alpha}(\nu,m) v_{\gamma}(-\nu,m) \hat{h}_{\alpha\gamma\sigma}
e_{n\sigma}(\nu,m) ,
\label{annums}
\end{eqnarray}
\begin{eqnarray}
e_{n\sigma}(\nu,m) = \tilde{g}_{L\sigma}(n-\nu)\tilde{g}_{L\sigma}(n) .
\label{ennum}
\end{eqnarray}
Here $\hat{h}_{\alpha\gamma\sigma} = \delta_{\alpha\gamma} 
+ \sigma(1-\delta_{\alpha\gamma})$, and we used a notation 
$\tilde{g}_{L\sigma}(n) = \tilde{g}_{L\sigma\sigma}(i\omega_{n})$ for
simplicity. 
 
In order to reduce these summations, we make use of an
asymptotic approximation~\cite{kake02,kake10}.  
\begin{eqnarray}
e_{n\sigma}(\nu, m) \sim 
\overline{q}_{\nu} \, \big( 
\tilde{g}_{L\sigma}(n-\nu) - \tilde{g}_{L\sigma}(n) \big)
\ ,
\label{easym}
\end{eqnarray}
where $\overline{q}_{\nu}=\beta/2\pi\nu i$.
The approximation is justified in the
high-frequency limit where $\tilde{g}_{L\sigma}(n)$ is written as
\begin{eqnarray}
\tilde{g}_{L\sigma}(n) = 
\dfrac{1}{i\omega_{n} - \epsilon^{0}_{L} + \mu - v_{L\sigma}(0)}
+ O\left(\dfrac{1}{(i\omega_{n})^{3}} \right) .
\label{gasym}
\end{eqnarray}
In the asymptotic approximation,
we obtain
\begin{eqnarray}
\hat{D}^{(n)}_{\{\alpha\gamma\}}(\nu, k, m) &=&
\sum^{n}_{l=0} \hat{D}^{(l)}_{\{\alpha_{1}\gamma_{1}
\cdots \alpha_{l}\gamma_{l}\}\uparrow}(\nu, k, m)\nonumber\\
& &\hspace*{-2mm}\times\hat{D}^{(n-l)}_{\{\alpha_{l+1}\gamma_{l+1}
\cdots \alpha_{n}\gamma_{n}\}\downarrow}(\nu, k, m) \ .
\label{dnag2}
\end{eqnarray}
Here we wrote the subscript at the r.h.s. (right-hand-side) explicitly 
to avoid confusion.  Note that the values of $\alpha_{i}$ and 
$\gamma_{i}$ are limited to $0$ or $z$ in the present approximation.
The spin-dependent quantities are given by~\cite{kake10}
\begin{eqnarray}
\hat{D}^{(l)}_{\{\alpha\gamma\}\sigma}(\nu, k, m) =
\Lambda^{(l)}_{\sigma}(\{\alpha\gamma\}) 
\frac{\overline{q}_{\nu}^{\,i}}{l\,!} B^{(l)}_{\sigma}(\nu, k, m)
\ ,
\label{dlags}
\end{eqnarray}
\begin{eqnarray}
\Lambda^{(l)}_{\sigma}(\{\alpha\gamma\}) = 
\begin{cases}
\ 1 & (\sigma=\uparrow) \\
(-1)^{l-n_{l}(\{\alpha\gamma\})} & (\sigma=\downarrow)
\end{cases}
\ ,
\label{lambda}
\end{eqnarray}
\begin{widetext}
\begin{eqnarray}
B^{(l)}_{\sigma}(\nu, k, m) = 
\Big[ \prod^{l-1}_{j=0} \tilde{g}_{L\sigma}(j\nu+k) \Big] 
+ \sum^{l-1}_{i=0} \dfrac{(-)^{l-i} l!}{i! (l-i)!}
\Big[ \prod^{i-1}_{j=-(l-i)} \tilde{g}_{L\sigma}(j\nu+k) \Big]
\Big[ 1 + \dfrac{l-i}{\overline{q}_{\nu}^{\,i}} 
\tilde{g}_{L\sigma}(i\nu+k) \Big] \ .
\label{bls}
\end{eqnarray}
\end{widetext}
Here $\hat{D}^{(0)}_{\{\alpha\gamma\}\sigma}(\nu, k, m) = 1$.
$n_{l}(\{\alpha\gamma\})$ is the number of $\{\alpha_{i}\gamma_{i}\}$ 
pairs such that $\alpha_{i}=\gamma_{i}$ among the $l$ pairs.
Equation (\ref{bls}) reduces to the
result of the zeroth asymptotic approximation in our paper 
I~\cite{kake02} when there is no orbital degeneracy.

In the actual applications we make use of the exact form up to a certain
order of expansion in $\overline{D}^{(m)}_{\nu}$ (see Eq. (\ref{dnukm2})), 
and for higher-order terms we adopt an asymptotic
form (\ref{dnag2}).  
In this way, we can take into account dynamical corrections
systematically starting from both sides of the weak-interaction limit 
and the high-temperature limit.

The coherent potential can be determined by the stationary condition
$\delta \mathcal{F}_{\rm CPA}/\delta \Sigma = 0$.  
This yields the dynamical CPA equation as~\cite{kake08} 
\begin{eqnarray}
\langle G_{L\sigma}(\mbox{\boldmath$\xi$}, i\omega_{n}) \rangle 
= F_{L\sigma}(i\omega_{n}) \ .
\label{dcpa3}
\end{eqnarray}
Here $\langle \ \rangle$ at the l.h.s. (left-hand-side) 
is a classical average taken with respect to the
effective potential $E_{\rm eff}(\mbox{\boldmath$\xi$})$.
The impurity Green function is given by 
\begin{eqnarray}
G_{L\sigma}(\mbox{\boldmath$\xi$}, i\omega_{l}) = 
\tilde{g}_{L\sigma\sigma}(i\omega_{l}) + 
\dfrac{\displaystyle \sum_{n} \sum_{\nu} 
\frac{\delta \overline{D}^{(n)}_{\nu}}
{\displaystyle \kappa_{L\sigma}(i\omega_{l})
\delta \Sigma_{L\sigma}(i\omega_{l})}}
{\displaystyle 1+ \sum_{n} \sum_{\nu} \overline{D}^{(n)}_{\nu}} \ .
\nonumber \hspace{-10mm}\\
\label{gimp}
\end{eqnarray}
The first term at the r.h.s. is the impurity 
Green function in the static approximation, which is given 
by Eq. (\ref{gst}). 
The second term is the dynamical corrections, and
$\kappa_{L\sigma}(i\omega_{l})= 1 - F_{L\sigma}(i\omega_{l})^{-2}
\delta F_{L\sigma}(i\omega_{l})/\delta \Sigma_{L\sigma}(i\omega_{l})$.

Solving the CPA equation (\ref{dcpa3}) self-consistently, we obtain 
the effective medium.  
The electron number on each orbital $L$ is then calculated from
\begin{eqnarray}
\langle \hat{n}_{L} \rangle = 
\dfrac{1}{\beta} \sum_{n\sigma} F_{L\sigma}(i\omega_{n}) \ .
\label{avnl}
\end{eqnarray}
The chemical potential $\mu$ is determined from the condition 
$n_{e} = \sum_{L} \langle \hat{n}_{L} \rangle$.
Here $n_{e}$ denotes the conduction electron number per atom.
The magnetic moment is given by
\begin{eqnarray}
\langle \hat{m}^{z}_{L} \rangle = 
\dfrac{1}{\beta} \sum_{n\sigma} \sigma F_{L\sigma}(i\omega_{n}) \ .
\label{avml}
\end{eqnarray}
In particular, the $l=2$ component of magnetic moment is expressed 
as
\begin{eqnarray}
\langle \hat{\boldsymbol{m}}_{l} \rangle = 
\langle \boldsymbol{\xi} \rangle \ . 
\label{avmd}
\end{eqnarray}
The above relation implies that the effective potential 
$E_{\rm eff}(\mbox{\boldmath$\xi$})$
is a potential energy for a local magnetic moment
$\mbox{\boldmath$\xi$}$.

In the numerical calculations, we took into account the dynamical
corrections up to the second order ($n \le 2$) exactly, 
and the higher-order terms
up to the fourth order within the asymptotic approximation.
Summation with respect to $\nu$ in Eqs. (\ref{edyn2})
and (\ref{gimp}) was taken up to $\nu = 100$ for $n=1$ and
$2$, and up to $\nu = 2$ for $n = 3, \ 4$.

When we solve the CPA equation (\ref{dcpa3}), we adopted a decoupling 
approximation to the thermal average of impurity Green 
function~\cite{kake81} for simplicity, 
{\it i.e.}, 
\begin{eqnarray}
\langle G_{L\sigma}(\xi_{z}, \xi^{2}_{\perp}, i\omega_{n}) \rangle &=& 
\sum_{q=\pm} \frac{1}{2}
\left( 1 + q \dfrac{\langle \xi_{z} \rangle}
{\sqrt{\langle \xi^{2}_{z} \rangle}} \right)\nonumber\\
& &\hspace{-3mm}\times G_{L\sigma}(q\sqrt{\langle \xi^{2}_{z} \rangle}, 
\langle \xi^{2}_{\perp} \rangle, i\omega_{n}) \ .
\label{gapprox}
\end{eqnarray}
Here we wrote the static exchange field $\boldsymbol{\xi}$ as 
$(\xi_{z}, \xi^{2}_{\perp})$ in order to make the decoupling 
approximation clearer.
The approximation is correct up to the second moment ({\it i.e.},
$\langle \xi^{2}_{\alpha} \rangle$) and arrows us to 
describe the thermal spin fluctuations in a simpler way.

On the other hand, we adopted a diagonal approximation~\cite{kirk70} 
to the coherent Green function at the r.h.s. of Eq. (\ref{dcpa3}). 
\begin{eqnarray}
F_{L\sigma}(n) = \int \dfrac{\rho^{\rm LDA}_{L}(\epsilon) d \epsilon}
{i\omega_{n} - \epsilon - \Sigma_{L\sigma}(i\omega_{n}) - 
\Delta\epsilon_{L}} \ .
\label{cohg2}
\end{eqnarray}
Here $\rho^{\rm LDA}_{L}(\epsilon)$ is the local density of states for the LDA
band calculation, and 
$\Delta \epsilon_{L} = (\epsilon_{L}-\epsilon^{0}_{L})\delta_{l2}$.
The approximation partly takes into account the effects of hybridization
between different $l$ blocks in the nonmagnetic state, but neglects the
effects via spin polarization.

The CPA equation (\ref{dcpa3}) 
with use of the decoupling approximation (\ref{gapprox})
yields an approximate solution to the full CPA equation.  
For the calculations of the single-particle densities of states, 
we adopted the following average $t$-matrix
approximation~\cite{korr58,ehren76} (ATA) after we solved 
Eq. (\ref{dcpa3}) with use of the decoupling approximation 
(\ref{gapprox}).  
\begin{eqnarray}
 \Sigma^{\rm ATA}_{L\sigma}(i\omega_{n}) =  \Sigma_{L\sigma}(i\omega_{n}) + 
\dfrac{\langle G_{L\sigma}(\xi_{z}, \xi^{2}_{\perp}, i\omega_{n})
\rangle -F_{L\sigma}(i\omega_{n})}
{\langle G_{L\sigma}(\xi_{z}, \xi^{2}_{\perp}, i\omega_{n}) \rangle
F_{L\sigma}(i\omega_{n})} \ .
\nonumber \hspace{-10mm}\\
\label{ata}
\end{eqnarray}
Here the coherent potential in the decoupling
approximation is used at the r.h.s., but the full average 
$\langle \ \rangle$ of the 
impurity Green function is taken.  The ATA is a one-shot correction to
the full CPA.

The coherent potential $\Sigma_{L\sigma}(z)$ on the real axis $z=\omega
+ i\delta$ is then calculated by using the Pad\'{e}
numerical analytic continuation method~\cite{vidberg77}. 
Here $\delta$ is an infinitesimal positive number.
The densities of states (DOS) as the single-particle excitations, 
$\rho_{L\sigma}(\omega)$ are 
calculated from the relation,
\begin{eqnarray} 
\rho_{L\sigma}(\omega) = - \frac{1}{\pi} \, {\rm Im} \, F_{L\sigma}(z) \ .
\label{dos}
\end{eqnarray}
%
%
%
%---------------------------------------------------------------------
\begin{figure}
\includegraphics[scale=0.75]{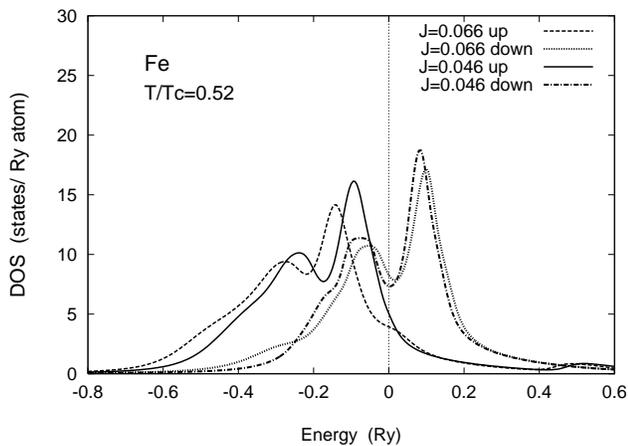}%
\caption{\label{figfedosd}
Spin-resolved $d$ densities of states (DOS) of Fe at $T/T_{\rm C}=0.52$
 for atomic $J$ (dashed curve for the up-spin DOS, dotted curve for 
the down-spin DOS) and screened $J$ (solid curve for the up-spin DOS, 
dot-dashed curve for the down-spin DOS). 
}
\end{figure}
%---------------------------------------------------------------------
%
%

\section{Numerical results}

In the numerical calculations, we adopted the lattice constants 
used by Andersen {\it et. al.}~\cite{ander94}, and 
performed the LDA calculations with use of the Barth-Hedin
exchange-correlation potential to make the TB-LMTO Hamiltonian (\ref{h0}).
For Fe and Ni, we adopted average Coulomb interaction parameters 
$\overline{U}$ and the average exchange interactions $\overline{J}$ 
used by Anisimov {\it et. al.}~\cite{anis97-2}, and for Co we adopted 
$\overline{U}$ obtained by Bandyopadhyay {\it et. al.}~\cite{bdyo89} 
and $\overline{J}$ obtained by the Hartree-Fock atomic
calculations~\cite{mann67} 
; $\overline{U}=0.169$ Ry and 
$\overline{J}=0.066$ Ry for Fe, $\overline{U}=0.245$ Ry and 
$\overline{J}=0.069$ Ry for fcc Co, and $\overline{U}=0.221$ Ry and 
$\overline{J}=0.066$ Ry for Ni.
The exchange interaction energies mentioned above are basically atomic
ones.  
As for the screened exchange interaction energy, we adopted 70\% 
values~\cite{arya06,miyake08}
of these atomic $\overline{J}$; $\overline{J}=0.046$ Ry for Fe, 
$\overline{J}=0.048$ Ry for fcc Co, and 
$\overline{J}=0.046$ Ry for Ni. 
The intra-orbital Coulomb interaction $U_{0}$, inter-orbital Coulomb 
interaction $U_{1}$, 
and the exchange interaction energy parameter $J$ were calculated from 
$\overline{U}$ and $\overline{J}$ as 
$U_{0} = \overline{U} + 8\overline{J}/5$, 
$U_{1} = \overline{U}-2\overline{J}/5$, and $J=\overline{J}$, 
using the relation $U_{0} = U_{1}+2J$.  

We present in Fig. 1 the $d$ densities of states (DOS) in the 
ferromagnetic Fe.  
In the case of atomic $J=0.066$ Ry, the DOS for up-spin
electrons consists of two peaks, the main peak due to e${}_{g}$
electrons at $\omega \approx -0.15$ Ry and the second peak due to
t${}_{2g}$ electrons at $\omega \approx -0.30$ Ry.  We also find small
humps at $\omega \approx -0.5$ Ry and 0.0 Ry.  These humps are caused by
both e${}_{g}$ and t${}_{2g}$ electrons.
The DOS for down-spin electrons also show the structure consisting of
the main peak at $\omega = 0.1$ Ry mainly due to e${}_{g}$ electrons
and the second peak at $\omega \approx -0.05$ Ry mainly due to 
t${}_{2g}$ electrons.  
We find a small hump at $\omega \approx -0.3$ Ry due to
t${}_{2g}$ electrons.

When we adopt the screened $J=0.046$ Ry, the peaks
of the up-spin DOS shift up towards the Fermi level by about 0.05 Ry, and
those of the down-spin DOS shift down.  It indicates the
reduction of the magnetization.  
The hump at $\omega \approx -0.5$ Ry shifts up and loses its weight,
while the hump at $\omega \approx -0.3$ Ry disappears. 
Accordingly the spectral weights of the two peaks are
enhanced.  The disappearance of the humps implies that the humps 
originates in the multiplet excitations due to strong $J$. 
The resulting total DOS are shown in Fig. 2.
The screened $J$ enhances the main peak and shifts it to
the Fermi level.  The DOS obtained by the screened $J$ 
seems to explain better the XPS data at room temperature~\cite{narm88}. 
%
%
%---------------------------------------------------------------------
\begin{figure}
\includegraphics[scale=0.75]{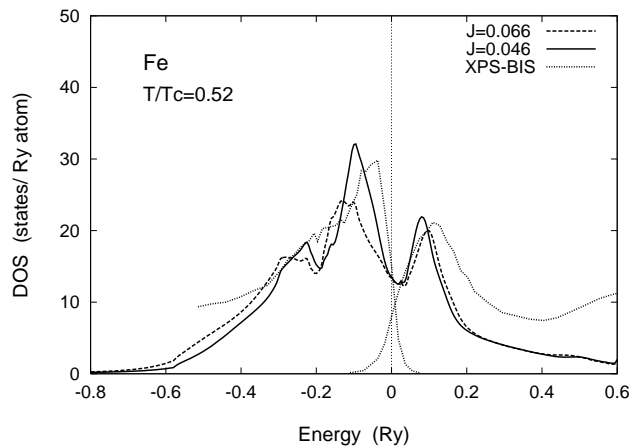}%
\caption{\label{figfedos}
Total DOS of Fe for atomic $J$ (dashed curve) and screened $J$ (solid curve)
at $T/T_{\rm C}=0.52$.
The XPS~\cite{narm88} and BIS~\cite{spei84} data at room temperature are 
shown by dotted curves.
}
\end{figure}
%---------------------------------------------------------------------
%
%

Solving the CPA equation (\ref{dcpa3}) self-consistently at each 
temperature, we obtain the magnetization vs temperature ($M-T$) curves 
for Fe as shown in Fig. 3.
The ground-state magnetizations $M(0)$ obtained by an extrapolation of the
$M-T$ curves are 2.58 $\mu_{\rm B}$ for $J=0.066$ Ry and 2.39 
$\mu_{\rm B}$ for $J = 0.046$ Ry.
The latter is in better agreement with experimental value~\cite{danan68} 
2.22 $\mu_{\rm B}$, though it is still somewhat overestimated.
The reduction of $M(0)$ with decreasing $J$ is explained by the fact
that the Hund-rule coupling $J$ which parallels the spins of electrons 
in each atom competes with the disordering due to electron hopping and 
thus $M(0)$ is not saturated. 
We also find the reduction of the amplitude of local moment 
$\langle \boldsymbol{m}^{2} \rangle^{1/2}$ by 0.066 $\mu_{\rm B}$ when the
screened $J$ is used, as seen in Fig. 3.
%
%
%---------------------------------------------------------------------
\begin{figure}
\includegraphics[scale=0.75]{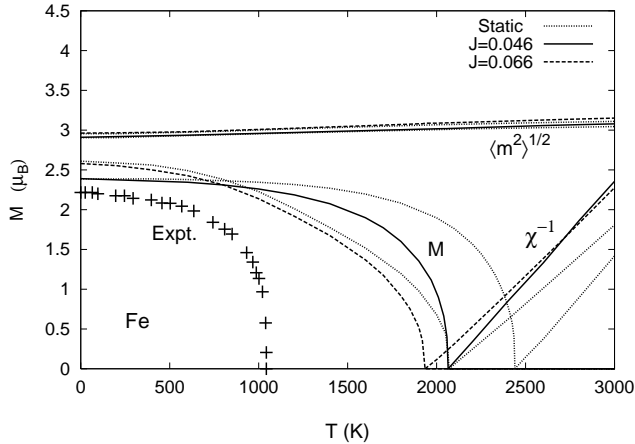}%
\caption{\label{figfemt}
Magnetization vs temperature ($M-T$) curves, inverse susceptibilities 
$\chi^{-1}$, and amplitudes of local moments 
$\langle \boldsymbol{m}^{2} \rangle^{1/2}$ of Fe for atomic $J$ (dashed curves) 
and screened $J$ (solid curves). Corresponding curves in the static
 approximation are drawn by dotted curves.  The curve with higher
 (lower) $T_{\rm C}$ corresponds to the screened (atomic) $J$.
Experimental data~\cite{potter34} of $M$ are shown by $+$.
}
\end{figure}
%---------------------------------------------------------------------
%
%

With increasing temperature, the magnetization for screened $J$
shows less temperature dependence at low temperatures and rapidly
decreases near $T_{\rm C}$.  Resulting $T_{\rm C}$ is higher than that
obtained by the atomic $J$.  Basically the Curie temperature $T_{\rm C}$
is determined by the ratio of the magnetic energy to the magnetic
entropy.  The enhancement of $T_{\rm C}$ with
reducing $J$ in Fe is attributed to the reduction of the magnetic entropy
with the collapse of the Hund rule coupling.
Further reduction of $J$ should decrease $T_{\rm C}$ because of the
reduction of magnetic energy.

Calculated susceptibilities of Fe follow the Curie-Weiss law as shown in
Fig. 3.  We find the effective Bohr magneton numbers $m_{\rm eff} = 3.0$
$\mu_{\rm B}$ for $J=0.066$ Ry and $m_{\rm eff} = 2.8$ $\mu_{\rm B}$ 
for $J=0.046$ Ry, which are compared with the experimental 
value~\cite{fallot44} 3.2 $\mu_{\rm B}$.  
The screened $J$ underestimates the effective Bohr magneton number.
We have also calculated the effective Bohr magneton number for fcc Fe at
high temperatures ($\sim 2000$ K), and obtained 
$m_{\rm eff}({\rm fcc}) = 4.2$ 
$\mu_{\rm B}$ for $J=0.066$ Ry and $m_{\rm eff}({\rm fcc}) = 2.6$ 
$\mu_{\rm B}$ for $J=0.046$ Ry after having included the volume effects 
on the susceptibilities.
These values should be compared with the experimental value~\cite{bar63} 
7.0 $\mu_{\rm B}$.
The results again show that the screened $J$ value underestimates
$m_{\rm eff}$.
Present results indicate that the high-temperature properties of Fe are
explained better by the atomic $J$, while the low-temperature ones are
explained by the screened $J$.
%
%
%---------------------------------------------------------------------
\begin{figure}
\includegraphics[scale=0.75]{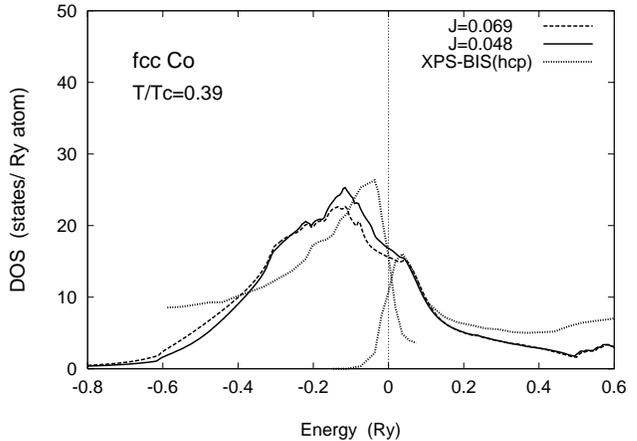}%
\caption{\label{figcodos}
Total DOS of fcc Co for atomic $J$ (dashed curve) and screened $J$ 
(solid curve) at $T/T_{\rm C}=0.39$.
The XPS~\cite{narm88} and BIS~\cite{spei84} data at room temperature 
for hcp Co are shown by dotted curves.
}
\end{figure}
%---------------------------------------------------------------------
%
%

In the case of fcc Co, the $d$ DOS for up-spin electrons shift up by
0.04 Ry and those for down-spin electrons shift down by 0.01 Ry when the
atomic $J$ is replaced by the screened $J$.  
The total DOS shifts towards the
Fermi level and the peak at $\omega = -0.1$ Ry is considerably enhanced
as shown in Fig. 4.

It should be noted that the XPS~\cite{narm88} and BIS~\cite{spei84} 
data are obtained for the hcp Co at room temperature, 
thus they are not able to be compared directly 
with the present results for fcc Co.  Nevertheless,
we find the peak position of the BIS data is in accordance with the hump
position in the DOS.  The band width of the DOS seems to be larger
than that expected from the XPS data.
%
%
%---------------------------------------------------------------------
\begin{figure}
\includegraphics[scale=0.75]{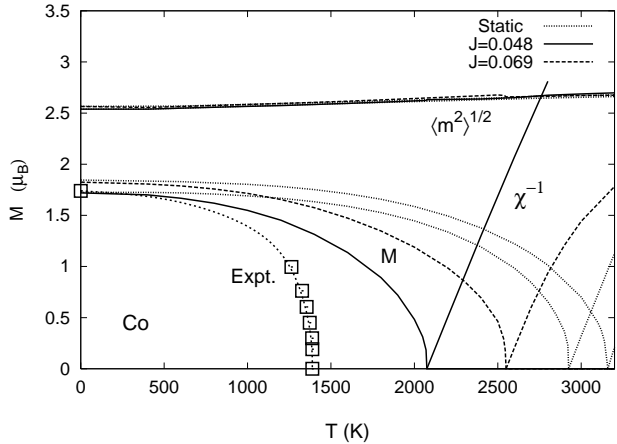}%
\caption{\label{figcomt}
Magnetization vs temperature ($M-T$) curves, inverse susceptibilities 
$\chi^{-1}$, and amplitudes of local moments 
$\langle \boldsymbol{m}^{2} \rangle^{1/2}$ of fcc Co for atomic $J$ 
(dashed curves) and screened $J$ (solid curves). Corresponding curves in 
the static approximation are drawn by dotted curves.  
The curve with higher (lower) $T_{\rm C}$ corresponds to the atomic
 (screened) $J$.  Experimental data~\cite{meyers51} 
of $M$ for fcc Co are shown by open squares and dashed line.
}
\end{figure}
%---------------------------------------------------------------------
%
%

We present the magnetization vs temperature curves of fcc Co in Fig. 5.  
The ground-state magnetizations obtained by an extrapolation are 1.82
$\mu_{\rm B}$ for the atomic $J=0.069$ Ry and 1.72 $\mu_{\rm B}$ for 
the screened $J = 0.048$ Ry, respectively; 
the latter obtained by the screened $J$ is in better 
agreement with the experimental one~\cite{bes70} (1.74 $\mu_{\rm B}$).
The amplitudes of local moments decrease by about 0.007 
$\mu_{\rm B}$ with the reduction of the Hund-rule coupling by 0.021 
Ry.

The magnetization for the screened $J$ decreases in the same
way as in the atomic $J$ with increasing temperature (see Fig. 5).  
The calculated $T_{\rm C}$ is 2075 K for the screened $J=0.048$ Ry, and
is smaller than $T_{\rm C}=2550$ K for atomic $J=0.069$ Ry.  
The former (latter) is 1.5 (1.8) times as large as 
the experimental value of fcc Co ($T_{\rm C}=1388$ K)~\cite{colvin65}.  
We also calculated the susceptibilities.  
The inverse susceptibilities follow the Curie-Weiss
law but show an upward convexity to some extent.
Calculated effective Bohr magneton numbers are 3.0 $\mu_{\rm B}$ in both
cases being in good agreement with the experimental 
value~\cite{fallot44} 3.15 $\mu_{\rm B}$.  
They hardly change due to the reduction of $J$.
These results indicate that effects of the Hund-rule coupling on the
local magnetic moment are less significant as compared with the case of
Fe because the number of $d$ holes is rather small (about 2.36).
%
%
%---------------------------------------------------------------------
\begin{figure}
\includegraphics[scale=0.75]{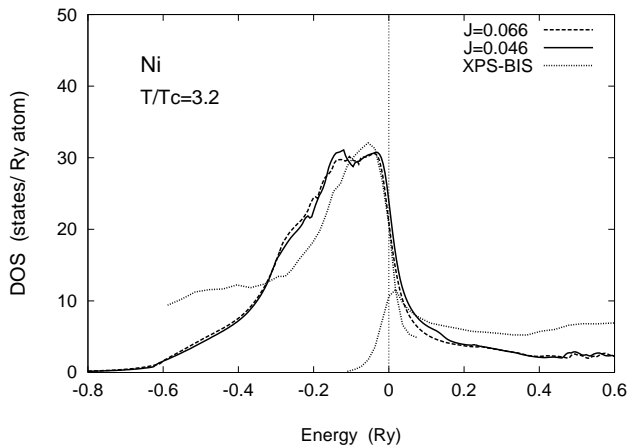}%
\caption{\label{fignidos}
Total DOS of Ni for atomic $J$ (dashed curve) and screened $J$ 
(solid curve) at $T/T_{\rm C}=3.2$.
The XPS~\cite{narm88} and BIS~\cite{spei84} data are shown by 
dotted curves.
}
\end{figure}
%---------------------------------------------------------------------
%
%

The DOS for Ni are presented in Fig. 6.  They are shown for
paramagnetic state because calculated $T_{\rm C}$ of Ni 
is lower than those in Fe and Co by a factor of 3 or 5, and the 
calculations of DOS at low temperatures below $T_{\rm C}$ are 
not easy in the present
method with use of the numerical analytic continuation. 
The paramagnetic DOS are not sensitive to the choice of $J$ since 
the Hund-rule coupling is not effective for 
the system with small number of $d$ holes ($\approx 1.32$).  
Calculated $d$-band widths seem to be somewhat larger
than those expected from the XPS data.
The same tendency was found also in fcc Co (see Fig. 4).
The results suggest that the higher-order dynamical corrections are
desired for the quantitative description of the DOS in these systems.

The magnetization vs temperature curves for Ni are presented in Fig. 7.
The ground-state magnetization obtained by an extrapolation is 0.64
$\mu_{\rm B}$ for atomic $J=0.066$ Ry and $0.46$ $\mu_{\rm B}$ for 
screened $J=0.046$ Ry. 
The experimental value~\cite{bes70} 0.62 $\mu_{\rm B}$ is in-between.
The amplitude of local moment for $J=0.066$ Ry is 1.93 $\mu_{\rm B}$,
for example, at $T=T_{\rm C}$.  The amplitude hardly changes even if 
the screened $J$ is adopted.  
This is because the Hund-rule coupling is
negligible for the system with small number of $d$ holes.
%
%
%---------------------------------------------------------------------
\begin{figure}
\includegraphics[scale=0.75]{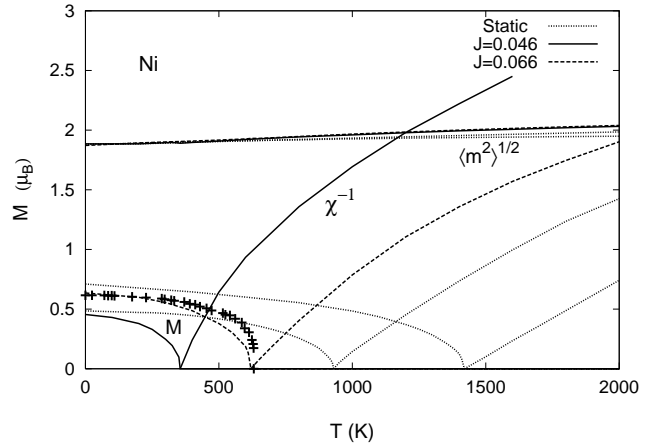}%
\caption{\label{fignimt}
Magnetization vs temperature ($M-T$) curves, inverse susceptibilities 
$\chi^{-1}$, and amplitudes of local moments 
$\langle \boldsymbol{m}^{2} \rangle^{1/2}$ of Ni for atomic $J$ (dashed
 curves) and screened $J$ (solid curves). Corresponding curves in 
the static approximation are drawn by dotted curves.  The curve with 
higher (lower) $T_{\rm C}$ corresponds to the atomic (screened) $J$. 
Experimental data~\cite{weiss26} of $M$ are shown by $+$.
}
\end{figure}
%---------------------------------------------------------------------
%
%

The Curie temperatures in Ni are strongly reduced by the dynamical 
effects as shown in Fig. 7.
We obtained $T_{\rm C} = 620$ K for the atomic $J$ (=0.066 Ry) and 355 K
for the screened $J$ (= 0.046 Ry).  The former is in good agreement with
the observed $T_{\rm C}=630$ K~\cite{noakes66}.
The screened $J$ underestimates the Curie temperature.  
The results indicate that the value of
$J$ is significant for the spin polarization and the inter-site 
ferromagnetic coupling in Ni.
In order to examine the ambiguity of the Coulomb interaction 
energy parameters, we performed the same
calculations with use of different sets of $(U, J)$: $(0.25, 0.073)$ Ry 
and (0.25, 0.051) Ry, where $U=0.25$ Ry was obtained by Bandyopadhyay 
{\it et. al.}~\cite{bdyo89} and $J=0.073$ Ry was taken from 
the Hartree-Fock atomic calculations~\cite{mann67}.
We obtained $T_{\rm C} = 680$ K for the former 
(, {\it i.e.}, atomic $J$) and 325 K for the latter 
(, {\it i.e.}, screened $J$); the results are not so sensitive to 
the choice of $U$. 

Calculated inverse susceptibilities are upwards convex,
being in agreement with the experimental data~\cite{fallot44} for Ni.  
Effective Bohr magneton
numbers calculated at $T \sim 2000$ K are 1.6 $\mu_{\rm B}$ for both the 
screened and the atomic $J$.  The results quantitatively agree with the
experimental value~\cite{fallot44} 1.6 $\mu_{\rm B}$.

\section{Summary}

We have calculated the magnetic properties of Fe, Co, and Ni at finite
temperatures on the basis of the first-principles dynamical CPA+HA, and
investigated the effects of screened exchange interaction 
energy $J$ on the magnetic properties in these ferromagnets.

When we apply the atomic $J$, we obtained from the magnetization vs
temperature curves the Curie
temperatures 1930K for Fe, 2550K for fcc Co, and 620K for Ni.
Although calculated $T_{\rm C}$ in Ni is close to the experimental 
value~\cite{noakes66} 630K, 
those in Fe and Co are overestimated by a factor of 1.8 
as compared with the experimental values, 
1040K (Fe)~\cite{arrott67} and 1388K (Co)~\cite{colvin65}, respectively.
In the present calculations using atomic $J$, 
the ground state magnetizations obtained by
an extrapolation are 2.58 $\mu_{\rm B}$ (Fe), 1.82 $\mu_{\rm B}$ (fcc Co),
and 0.64 $\mu_{\rm B}$ (Ni).  Calculated magnetizations for Fe and fcc
Co are considerably larger than the
experimental values, 2.22 $\mu_{\rm B}$ (Fe)~\cite{danan68} and 1.74 
$\mu_{\rm B}$ (Co)~\cite{bes70}, while that for Ni agrees well with the
experiment (0.62 $\mu_{\rm B}$).  
The results indicate that the atomic $J$ overestimate the
ferromagnetism of Fe and fcc Co at low temperatures.  
We have calculated the paramagnetic susceptibilities.  Calculated
effective Bohr magneton numbers, 3.0 $\mu_{\rm B}$ (Fe), 3.0
$\mu_{\rm B}$ (Co), and 1.6 $\mu_{\rm B}$ (Ni),
quantitatively explain the experimental data.

The reduction of the exchange interaction energy by 30\% due to 
screening by {\it sp} 
electrons weakens the Hund-rule coupling which builds up the local
magnetic moment on each atom.  We found that it reduces the ground-state
magnetization $M(0)$ and the amplitudes of local moments in Fe and Co, 
especially the ground-state magnetizations 2.39 $\mu_{\rm B}$ (Fe) and
1.72 $\mu_{\rm B}$ (fcc Co) are in good agreement with the experiments.
The screened $J$ reduces the exchange splitting between 
the up and down DOS.  The reduction of the splitting shifts the main 
peak in the spin-resolved DOS to the Fermi level and 
enhances the peak in case of Fe and Co.  
The screened $J$ reduces the humps in the DOS below 
$T_{\rm C}$ which originate
in the multiplet-type of excitations associated with $J$.
The screened $J$ also weakens the inter-site magnetic couplings.  It
reduces the magnetization in Ni, where the on-site Hund-rule coupling is
not significant because of a small number of $d$ holes.

With reducing $J$, the magnetic entropy and the magnetic energy decrease
in general.  We found that the 30\% screened $J$ enhances the
Curie temperature $T_{\rm C}$ by 140 K in Fe because of the reduction of
the magnetic entropy, while it reduces $T_{\rm C}$ in fcc Co by 480 K 
because of the reduction of the magnetic energy.
The Curie temperatures of Fe and Co are overestimated by a factor of
$1.5 \sim 2.0$ irrespective of the choice of $J$ in the present
calculations.  
It should be attiributed to the magnetic short range order which is 
not taken into account in the present theory.
In the case of the susceptibilities, the screened $J$ excessively 
reduces the effective Bohr magneton number in Fe 
where the Hund-rule coupling builds up the on-site magnetic moment. 

By comparing both results of atomic $J$ and screened $J$ with
the experimental data, we found that the screened $J$ explains
better the ground-state magnetization $M(0)$ and the DOS at low 
temperatures in case of Fe and Co.  
But it yields the underestimate of $M(0)$ in Ni.
The atomic $J$ seems to be better for the explanation of $M(0)$ in Ni.
Concerning the finite-temperature properties, we found that the screened
$J$ improves the Curie temperature in Co, while the atomic $J$ yields
better $T_{\rm C}$ in case of Fe and Ni.  
The effective Bohr magneton numbers are quantitatively explained 
by the atomic $J$.
The screened $J$ underestimates $m_{\rm eff}$ for both the bcc and fcc Fe.

The above-mentioned results indicate that only for fcc Co the screened
$J$ seems to improve both the low- and the high-temperature properties.
In the case of Fe, the high-temperature properties seem to be explained
better by the atomic $J$, while the low-temperature ones by the screened
$J$.  In Ni, both the low- and the high-temperature
properties are explained better by the atomic $J$.

Although we have to improve further the theory of dynamical 
CPA+HA at low-temperatures 
in order to obtain more solid conclusion, the present results 
obtained in the high-temperature region should be reasonable because
the theory is an approach from the high-temperature limit
and the magnetic short range order is not significant at high temperatures.  
The results for the effective Bohr magneton numbers 
indicate that one has to take into account the temperature dependence 
of $J$ for the system like Fe where the Hund-rule coupling builds a
large local magnetic moment but competes with the kinetic energy of $d$ 
electrons.  It is plausible that spin fluctuations at finite
temperatures break the screening of $J$.
The present results also suggest that the 30\% screening of $J$ is 
an overestimate for Ni.  The range of applications of
screened $J$ and its validity have to be examined further in the 
future investigations with use of more advanced theories.

\begin{acknowledgments}

The present work is supported by Grant-in-Aid for Scientific Research
(22540395). 
Numerical calculations have been partly carried out with use of the
Hitachi SR11000 in the Supercomputer Center, Institute of Solid State
Physics, University of Tokyo.

\end{acknowledgments}


\end{document}